\documentstyle[sprocl,epsf]{article}
\scrollmode
\newcommand{\dspexp}[1]{\mbox{\rm e\raisebox{2ex}{$\displaystyle{#1}$}}}

\begin{document}
\title{Flux-Bubble   Models and  Mesonic   Molecules} 

\author{M.M. Boyce\footnote{Speaker.}, J.  Treurniet, and P.J.S. Watson}

\address{Ottawa-Carleton Institute  for  Physics, Carleton   University,
Ottawa, Ontario,   Canada,  K1S-5B6} 

\maketitle 

\abstracts{It has   been   shown that the  string-flip   potential model
reproduces most  of the  bulk properties  of  nuclear matter,  with  the
exception of nuclear binding.  Furthermore,  it was postulated that this
model, with  the inclusion of   the colour-hyperfine interaction, should
produce   binding.  In  some  recent  work  a  modified   version of the
string-flip potential model was developed, called the flux-bubble model,
which would allow for the addition of perturbative QCD interactions.  In
attempts  to  construct a  simple $q\bar  q$   nucleon system using  the
flux-bubble  model  (which only   included  colour-Coulomb interactions)
difficulties arose with trying to construct a many-body variational wave
function that would  take into account  the locality of  the flux-bubble
interactions.  In this talk we consider a toy system, a mesonic molecule
in  order  to understand   these difficulties.  {\it   En  route}, a new
variational wave function is proposed that may have a significant enough
impact on the old string-flip potential model results that the inclusion
of perturbative effects may not be needed.}

\section{Introduction}

  For the  past 30 years  several  attempts have been  made, with little
success, to describe nuclear matter in  terms of its constituent quarks.
The main   difficulty is the non-perturbative  nature  of QCD.  The only
rigorous method for handling multi-quark systems to date is lattice QCD.
However, this is very computationally intensive  and given the magnitude
of the problem it appears unlikely to be useful in the  near future.  As
a result, more phenomenological means must be considered.

  A good phenomenological model  should be able  to reproduce, at  least
qualitatively, all the  overall  bulk properties of nuclear   matter, in
particular:

\mbox{}

\noindent
\begin{tabular}{@{$\;\;\bullet\:\:$}l}
nucleon gas at low densities with no power-law van der Waals forces\\
nucleon binding at higher densities\\
nucleon   swelling  and saturation  of   nuclear  forces with increasing
density\\
quark gas at extremely high densities.
\end{tabular}

\mbox{}

\noindent
There  are many models  that  attempt to reproduce  these properties but
none of them does so  completely.  In this   talk, only the  string-flip
potential   model$\,$\cite{kn:Boyce,kn:HorowitzI,kn:Watson}     will  be
considered.  This model  appears to be  promising  because it reproduces
most of  the aforementioned  properties with  the  exception  of nucleon
binding.  In  an attempt to   find  the root  of   this  problem, a  toy
mesonic-molecular  system   will be   used   to study  the  inclusion of
perturbative         interactions        and     variational        wave
functions.\cite{kn:thesisb}

  This  proceeding  is     essentially  part  II   of     our MRST   '94
talk.\cite{kn:BoyceA} For   completeness sake, a  brief overview  of the
previous proceedings will be given, followed by our most recent work.

\section{From String-Flips to Flux-Bubbles}

  In  this a   section, a  brief   outline of  the  generic  string-flip
potential model is presented, followed by an overview of the flux-bubble
model.

\subsection{The String-Flip Potential Model}
\label{sc:foo}

The string-flip  potential model postulates   how flux-tubes should form
between quarks at zero temperature, based on  input from lattice QCD and
experiment.  An adiabatic  assumption is made, in  which the quarks move
slowly enough for their fields to reconfigure themselves, such that the
overall potential energy is minimized: {\it i.e.},
\begin{equation}
V = \sum_{\min\{q_m\ldots q_n\}}{\bf v}({\vec{\rm r}}_m\ldots 
    {\vec{\rm r}}_n)
  = \left\{\mbox{\setlength{\unitlength}{1cm}\begin{picture}(2.75,1.25)
    \put(0.0625,-1.25){\mbox{\epsfxsize=3cm\epsffile{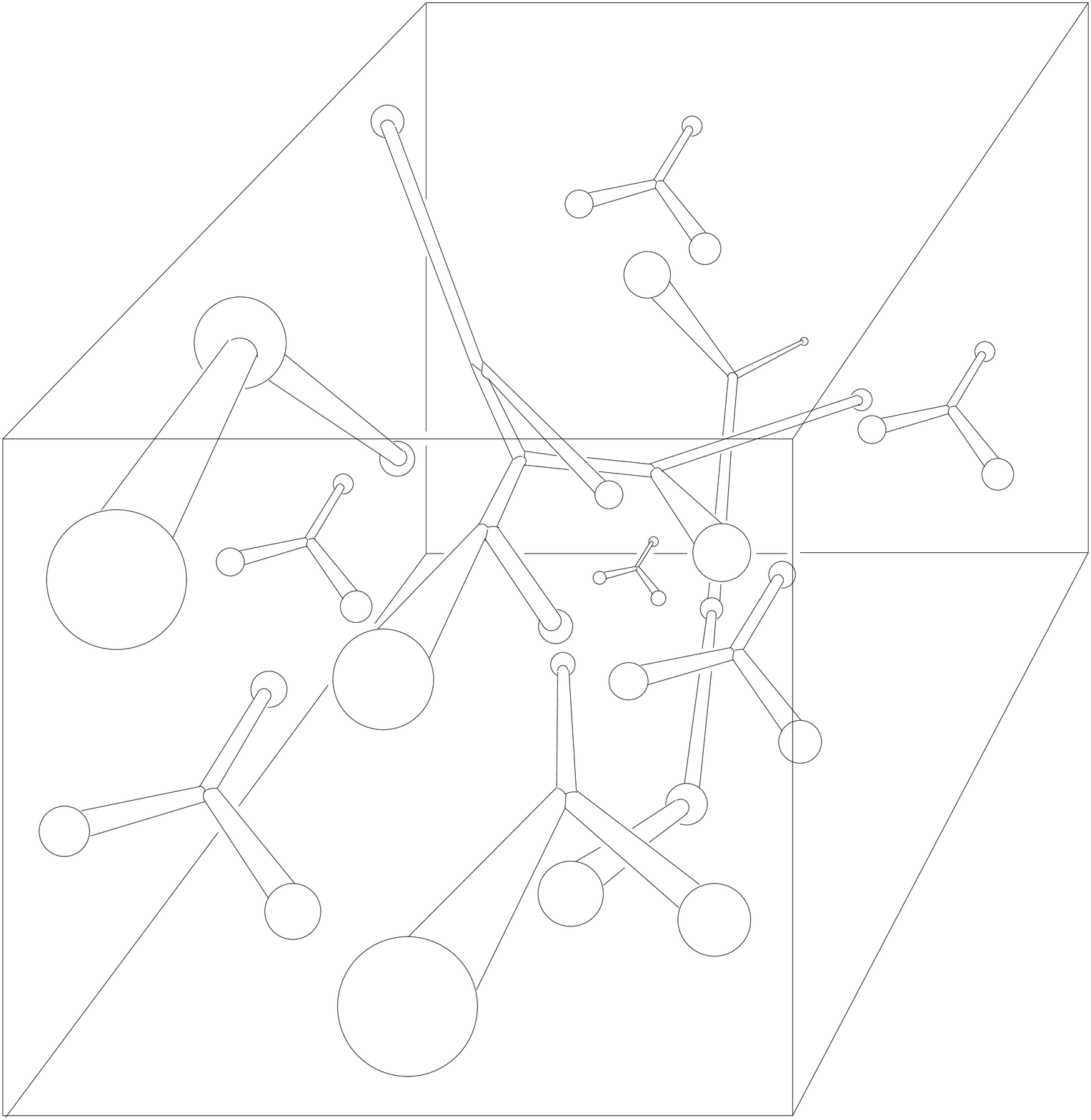}}}
    \end{picture}}\right\}\,.
    \label{eq:lincou}
\end{equation}
In  Eq.~\ref{eq:lincou},  the  sum is  over   all gauge invariant  sets,
$\{q_m\ldots q_n\}$, of $N_q$ quarks, $\{q_1\ldots  q_{N_q}\}$, in a box
of  side L  with periodic boundary   conditions (to  simulate continuous
nuclear matter).

  The physics, or model input, is  contained within the many-body terms,
${\bf  v}({\vec{\rm r}}_m\ldots  {\vec{\rm r}}_n)$,  in  $V$, which,  in
general,  can be  quite  complex  and  virtually  impossible to compute.
Ironically, the  form of ${\bf  v}$ is quite  simple:  the total minimal
length,   $\ell$, of flux-tubing (or string)   connecting a given set of
quarks times  the string  tension, $\sigma$,  ${\bf v}\sim\sigma\ell\,$.
Over the past  couple of years, many simplifications  of ${\bf  v}$ have
been assumed  in  the form  of  two-body, three-body,  and  long complex
chains of   quarks, using  $SU_c(2)$  and   $SU_c(3)$ with  and  without
``moving-colour.''$\,$\footnote{Typically,   a    quark  is  assigned  a
``fixed-colour''  which doesn't change  throughout  the evolution of the
gas.   Conversely, ``moving-colour,'' is  when the quarks are allowed to
change  colour.}$^,$\cite{kn:Boyce,kn:HorowitzI,kn:Watson}  All of these
models yield similar results for nuclear matter  --- no nuclear binding.
A general synopsis of these models can be  found in the MRST $^\prime$94
proceedings.\cite{kn:BoyceA} For simplicity, we shall concentrate on the
two-body quark-antiquark  form, ${\bf v}\sim\sigma r_{q\bar  q}\,$, with
$SU_c(2)$.

\subsection{The Flux-Bubble Model}
  
  To  date,  string-flip potential  models assumes only non-perturbative
interactions:  {\it  cf.}, ${\bf  v} \sim  \sigma\,  r -  \alpha_s/r\,$.
Flub-bubble models    are  a recent   attempt  to  include  perturbative
interactions.  In  particular, they  are a  derivative of  the following
two-body potential$\,$\cite{kn:BoyceA}
\begin{equation}
{\bf v}_{ij}=\sigma(r_{ij}-r_0)\theta(r_{ij}-r_0)+
\alpha_s\lambda_{ij}\left(\frac{1}{\rm r_{ij}}-\frac{1}{{\rm r}_0}\right)
\theta(r_0-r_{ij})\,,
\end{equation}
where $\lambda_{ij}$ is a  model dependent colour  factor, which is free
from Van der Waals forces.

   A ``flux-bubble'' is  formed when  two or  more ends of  neighboring
flux-tubes come  within ${\rm r}_0$  of each other, by inserting virtual
$q\bar  q$-pairs  on the  boundary   of  this  region,  thus  leaving a
perturbative bubble with non-perturbative flux-tubes emanating from it.
\begin{center}
\mbox{\setlength{\unitlength}{1mm}
\begin{picture}(110,57)
\put(-4,41){     \setlength{\unitlength}{1mm}     \begin{picture}(90,30)
   \put(0,0){\begin{minipage}{7.5cm}{\it  E.g}:{ \footnotesize  Consider
   configuration (a)  of quarks, with r  $>{\rm r}_0$,  about to move to
   (b), {\it s.t.},  two of them  are within  r $<{\rm r}_0$.   Then the
   procedure is to draw a bubble of ${\rm r}_0$ away  from the two, (b),
   and to cut the  flux-tubes at the  boundary and insert virtual $q\bar
   q$ pairs,  (c).  Once the  potential is computed the configuration is
   restored     to    (b)      before    the       next   move        is
   made.}\label{fig:fluxbubble}\end{minipage}}\end{picture}}
   \put(0,1){\mbox{\epsfxsize=11cm \epsffile{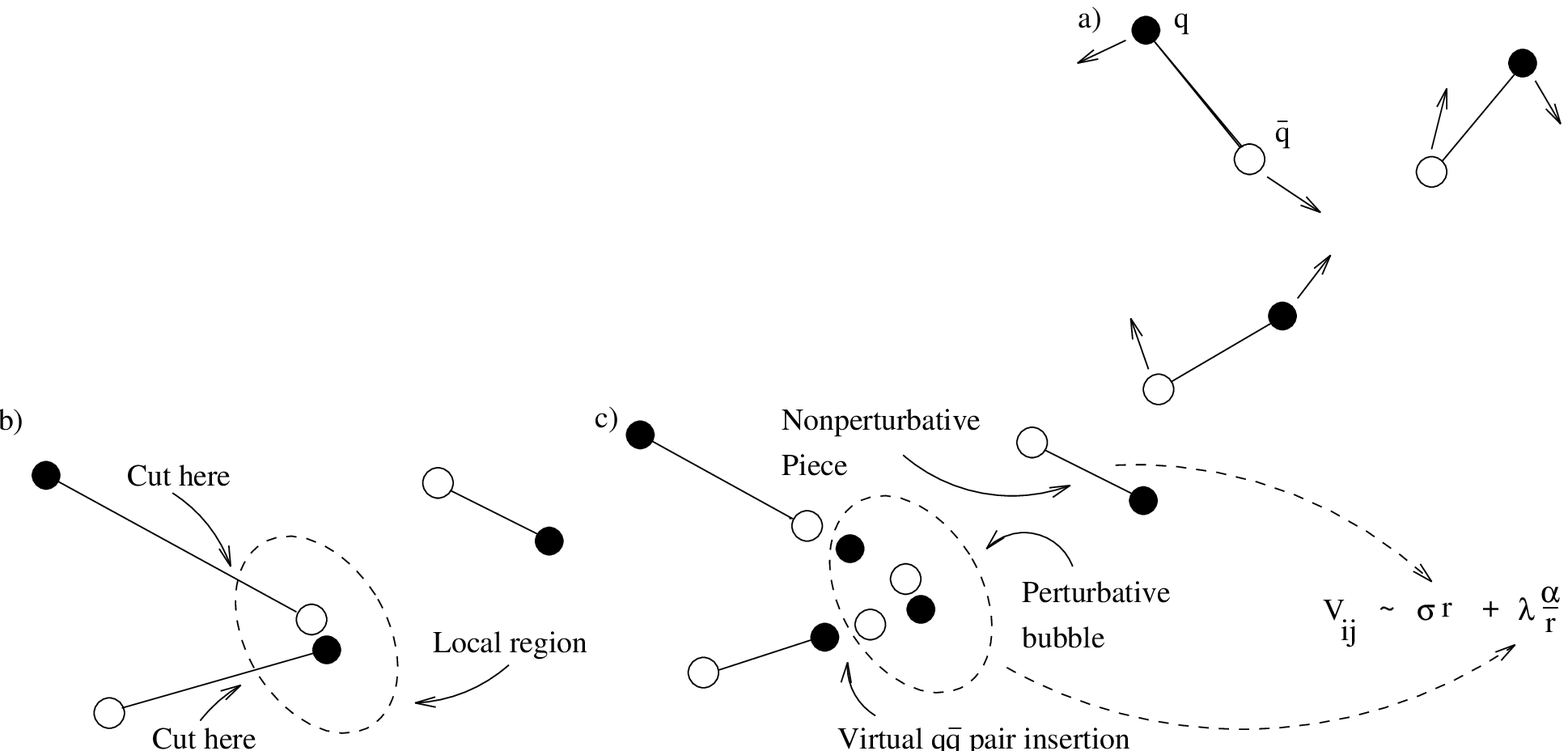}}}
\end{picture}}
\end{center}
Note  the  insertion   of  the  virtual   $q\bar  q$  pairs allows   the
construction of colourless objects. These  are solely used  as a tool to
calculate the overall length of the flux-tube correctly, and not used in
computing  the Coulomb term,  however, as  the field  energy is  already
taken into  account  by  the ``real''  quarks  inside  the   bubbles. In
general,  once the bubbles have  been determined, the flux-tubes must be
reconfigured in order to minimize the linear part of the potential.

  Therefore,  the most  general  $q\bar q$-flux-bubble   potential, with
$SU(2)$-colour, is given by ({\it cf.}, Eq.~\ref{eq:lincou})
\begin{equation}
V=\sigma\sum_{\min\{q\bar q\}}
      (r_{q\bar q}-r_0)\,\theta(r_{q\bar q}-r_0)+
      \alpha_s
      \sum_{i<j}\lambda_{ij}
      \left(\frac{1}{r_{ij}}-\frac{1}{r_0}\right)
      \theta(r_0-r_{ij})
\,,\label{eq:stringyb}
\end{equation}
where $\lambda_{ij}\,$ equals  $-3/4$  for unlike-quarks and   $1/4$ for
like-quarks.  The first    term represents the  string-flip, or  linear,
potential model, while second  term contains Coulomb corrections to  the
this potential  ({\it  i.e.}, linear-plus-Coulomb  potential model) plus
new   terms   describing  interactions    between   the  ends   of   the
flux-tubes.\footnote{It is conceivable,  that by extending the  model to
include interactions along the length of the flux-tubes, could lead to a
model for pion-liquid crystals.\cite{kn:Migdal}}

\subsection{Implementation Problems}

  The actual implementation  of the flux-bubble  potential turned out to
be  quite  straightforward.  However,  trying  to find  a good many-body
variational wave function proved to be  insurmountable. The main problem
was that each trial wave function required vast amounts of calculations
by hand,  writing  computer code, and lots   of computer  time  (running
several  machines   in  tandem).   Human  CPU   time ($\tau_{\mbox{\tiny
human}}\sim{\cal O}(1)wk$) aside,   using  a  simple  mesh  minimization
scheme  requires   ${\cal  O}(M^p/m)$ of  CPU  time  ($\tau_{\mbox{\tiny
CPU}}$), where $M$ is  the mesh size, $p$  is the number  of parameters,
and $m$ is the number of machines.

  A string-flip  potential model, for 7  triplets of quarks using fixed
$SU(2)$-colour, with a 3-parameter wave function,
\begin{equation}
\Psi_{\alpha\beta}=
\underbrace{{\rm e}^{-{\displaystyle \sum_{\min\{q\bar q\}}
   (\beta r_{q\bar{q}})^\alpha}
}\mbox{\rule[-3ex]{0em}{6ex}\hspace*{-2ex}}}_{\mbox{
\footnotesize Correlation $\Leftrightarrow$ $\beta$}}
\mbox{\hspace*{1ex}}
\underbrace{\prod_{\mbox{\tiny colour}}|\Phi(r_{p_k})|}_{\mbox{
\footnotesize Slater $\Leftrightarrow$ $\rho$}}\,,
\label{eq:watold}
\end{equation}
with parameters $\rho$ (density),  $\beta$ (inverse correlation length),
and $\alpha$, would  consume  about 5 weeks  of  CPU time  on our 8-node
farm, here at Carleton, using a crude $10^3$  mesh.  However, because of
certain assumptions  ($\alpha(\rho) \approx \alpha(\rho=0)$) and scaling
tricks  ($(\beta,\rho^{1/3}) \;  \longrightarrow  \;   \zeta(\theta)  \,
(\cos\theta,\sin\theta)$),   the CPU   time   can be   reduced  to 8hrs.
Unfortunately, because of the local nature  of the flux-bubble model, we
cannot use  scaling   tricks.    So,   in  general, we    would  require
$\tau_{\mbox{\tiny CPU}}\sim (10^{p-1}/21)wks$!
   
  Thus, a way of   checking  different wave functions  and  minimization
schemes which do not consume large amounts of CPU time is desirable.  In
particular, a mini-laboratory is needed in  which various aspects of the
string-flip and flux-bubble potential   models, from wave functions   to
minimization schemes, can be investigated,  without being too  concerned
about $CPU$  overhead.\footnote{In fact,   it was this   mini-laboratory
which enabled us to developed a  distributed minimization algorithm with
$\tau_{\mbox{\tiny CPU}}\sim{\cal O}(p/m)$.\cite{kn:thesisb}}

\section{Mesonic Molecules}

  An interesting  place that might   make  a good  mini-laboratory is  a
mesonic-molecule,\cite{kn:Weinstein}   $Q_2$, consisting     of      two
heavy-quarks, $Q$, and two relatively  light antiquarks, ${\bar q}$. The
heavy-quarks are assumed to be massive enough  so as not to be disturbed
by the    motion of  the  light-antiquarks: {\it   i.e.}, the  adiabatic
approximation.  By varying the distance, $R$, between the heavy-quarks a
mesonic-molecular potential, $U(R)$, can be computed.

  The Schr\"odinger   equation that  describes the  effective potential,
$U(R)$, is given by$\,$\cite{kn:Schiff}
\begin{equation}
\left(\frac{1}{2m_q}
\sum_{\bar q}\vec\nabla^2_{\bar q}+V\right)\Psi=U(R)\Psi\;\;\;\;
\mbox{\setlength{\unitlength}{1cm}\begin{picture}(4,0)
\put(1,-0.625){\mbox{\epsfxsize=3cm \epsffile{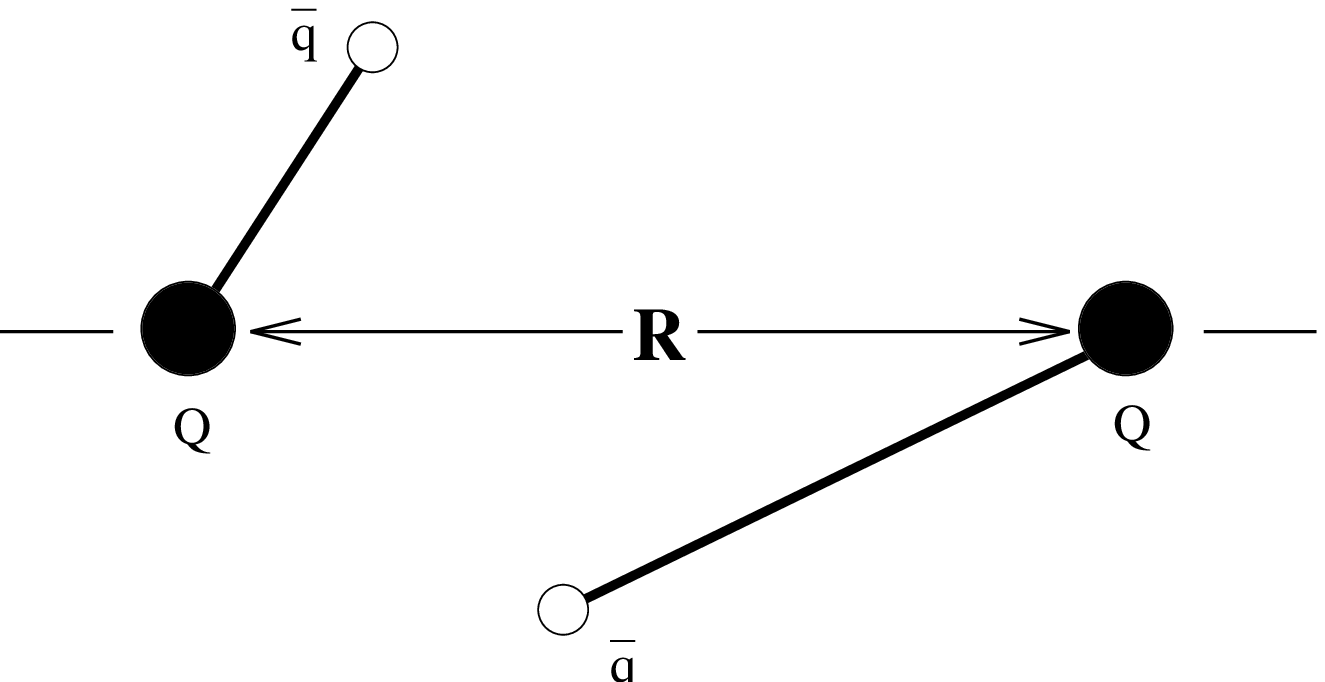}}}
\end{picture}}
\label{eq:Schiff}
\end{equation}
where $m_q=330\,MeV$.  The  potential,   $V$, describes   the  many-body
nature of the four quark system and is therefore  model dependent.  This
equation can be solved variationally for $\bar U(R)$, at fixed values of
$R$, by guessing the form the wave function, $\Psi$, and then minimizing
$\bar U=\bar T + \bar V$ with respect to the parameters in $\Psi$.

  Figs.~2.a,  shows   the results  for linear,  linear-plus-coulomb, and
flux-bubble   potential   models with  fixed-colour,     using the   old
string-flip variational   wave  function,\footnote{Figs.~2.a, also shows
results for the semi-relativistic Dirac-Hartree-Fock equation,
\begin{equation}
\sum_{\min\{Q\bar q\}}[\nabla^2_{\bar q}+(m_q+\sigma r_{Q\bar q})^2]\Psi
=U(R)^2\Psi\,
\label{fig:oneoverrsq}
\end{equation}
({\it cf.}, \ref{eq:Schiff}), for $m_q$ = 0 and 330 MeV.}
\begin{equation}
\Psi_{\alpha\beta}=
{\rm e}^{
      -{\displaystyle \sum_{\min\{Q\bar q\}}
      (\beta r_{Q\bar{q}})^\alpha}
      }\,,\label{eq:watwav}
\end{equation}
with parameters $\alpha\,$\footnote{Analysis  was  also done for   fixed
values of $\alpha\,$, at 2 and 1.74, which shows a 1\% variation in well
depth, suggesting   past assumptions about   $\alpha$  are valid.}   and
$\beta\,$.   These  models   yields an average    well depth  of  ${\cal
O}(3)MeV$, which is not enough to bind the $Q_2$ molecule. This suggests
that none of these models will  lead to nuclear binding,\footnote{We are
not surprised  that the flux-bubble model  yields very little change, as
the it was postulated   that the colour-hyperfine interactions are  more
important.\cite{kn:Boyce}} or  that perhaps something  is wrong with the
wave function itself.
\begin{center}
\mbox{ \setlength{\unitlength}{1mm}
\begin{picture}(110,130)
\put(-2,19){\setlength{\unitlength}{1mm}
\begin{picture}(110,110)
\put(10,105){\footnotesize (a)}
\put(65,105){\footnotesize (b)}
\put(10,50 ){\footnotesize (c)}
\put(65,50 ){\footnotesize (d)}
\put(0 ,55){\mbox{\epsfxsize=6cm\epsffile{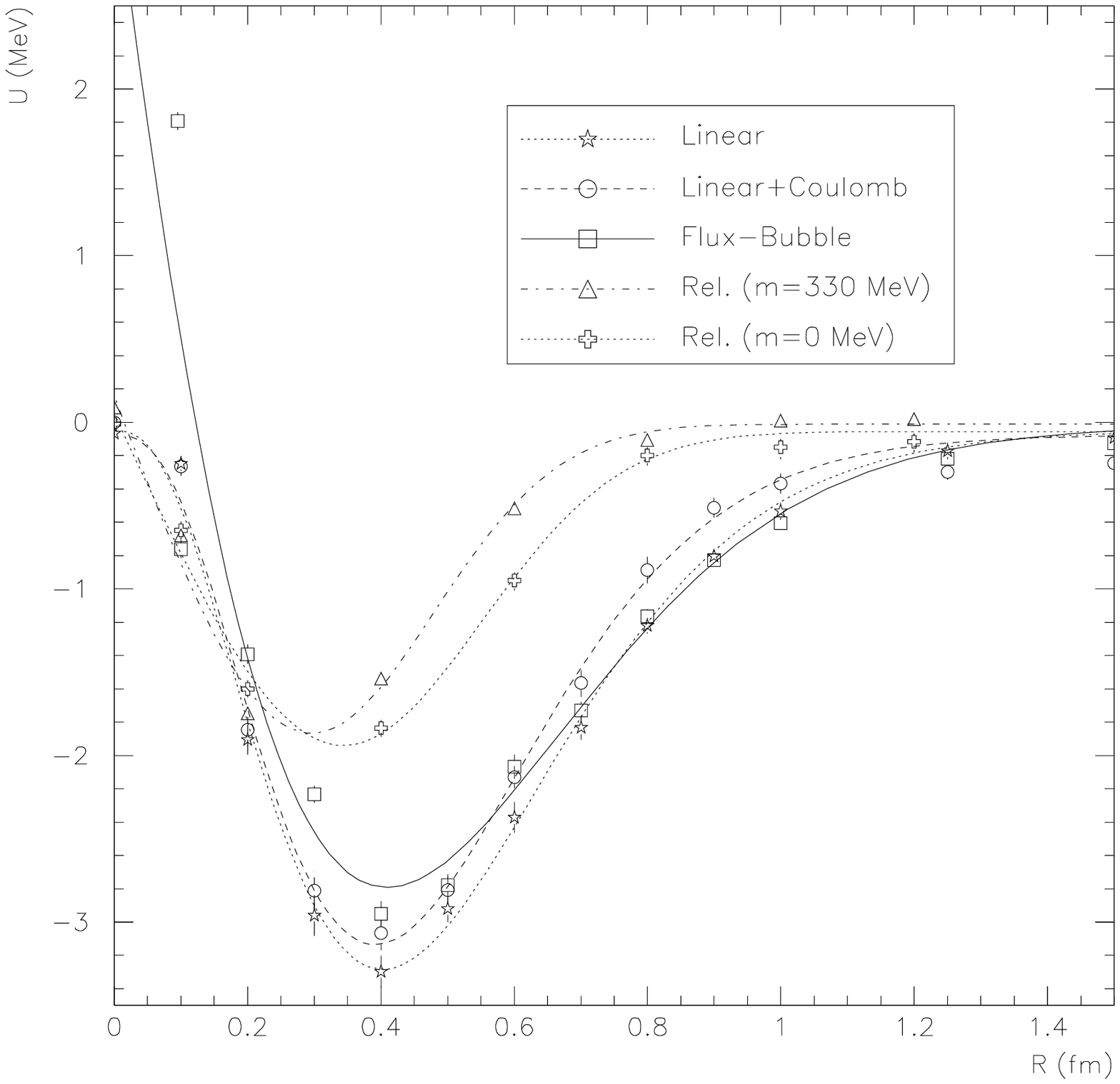}}}
\put(55,55){\mbox{\epsfxsize=6cm\epsffile{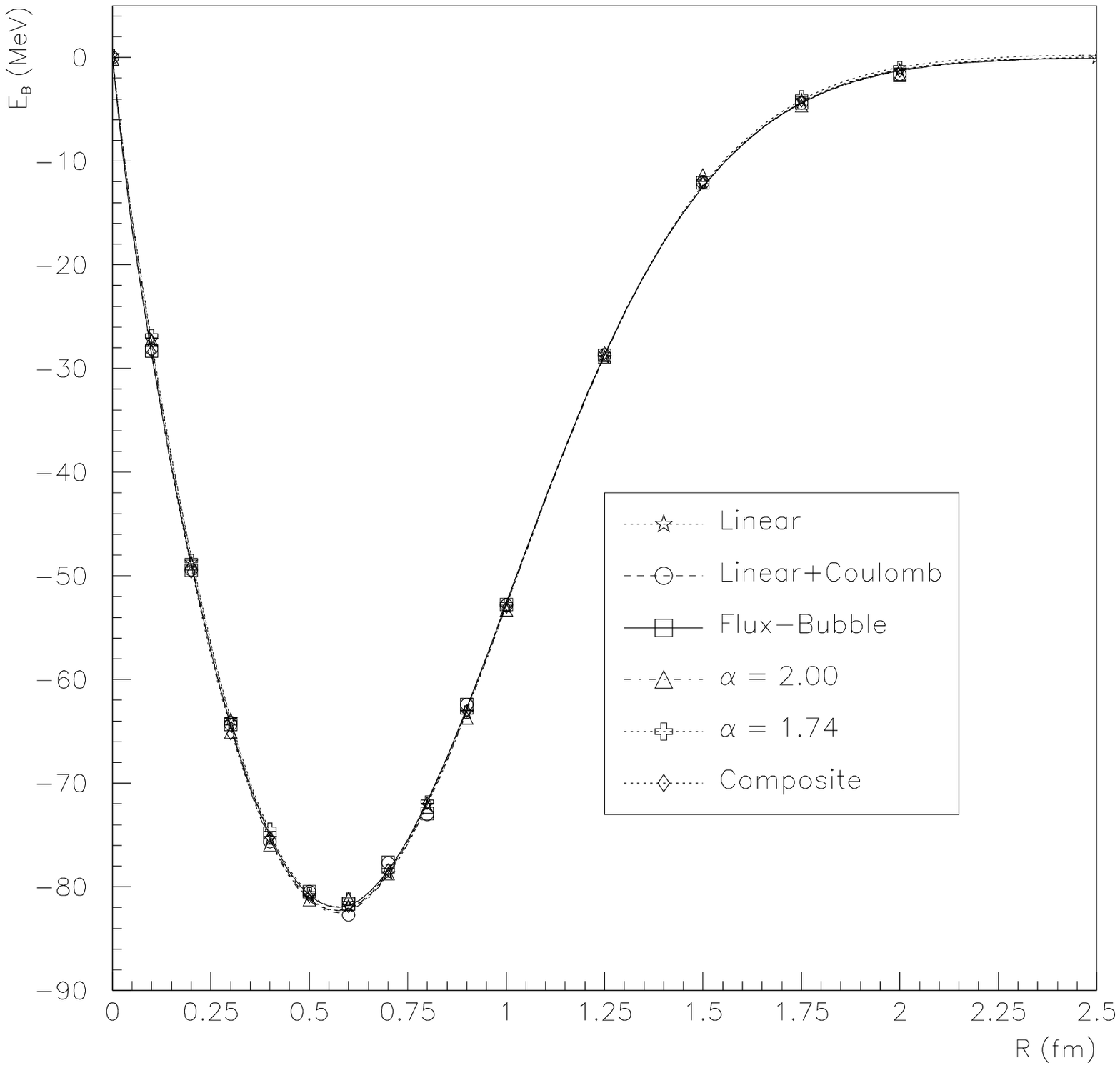}}}
\put(0 ,0 ){\mbox{\epsfxsize=6cm\epsffile{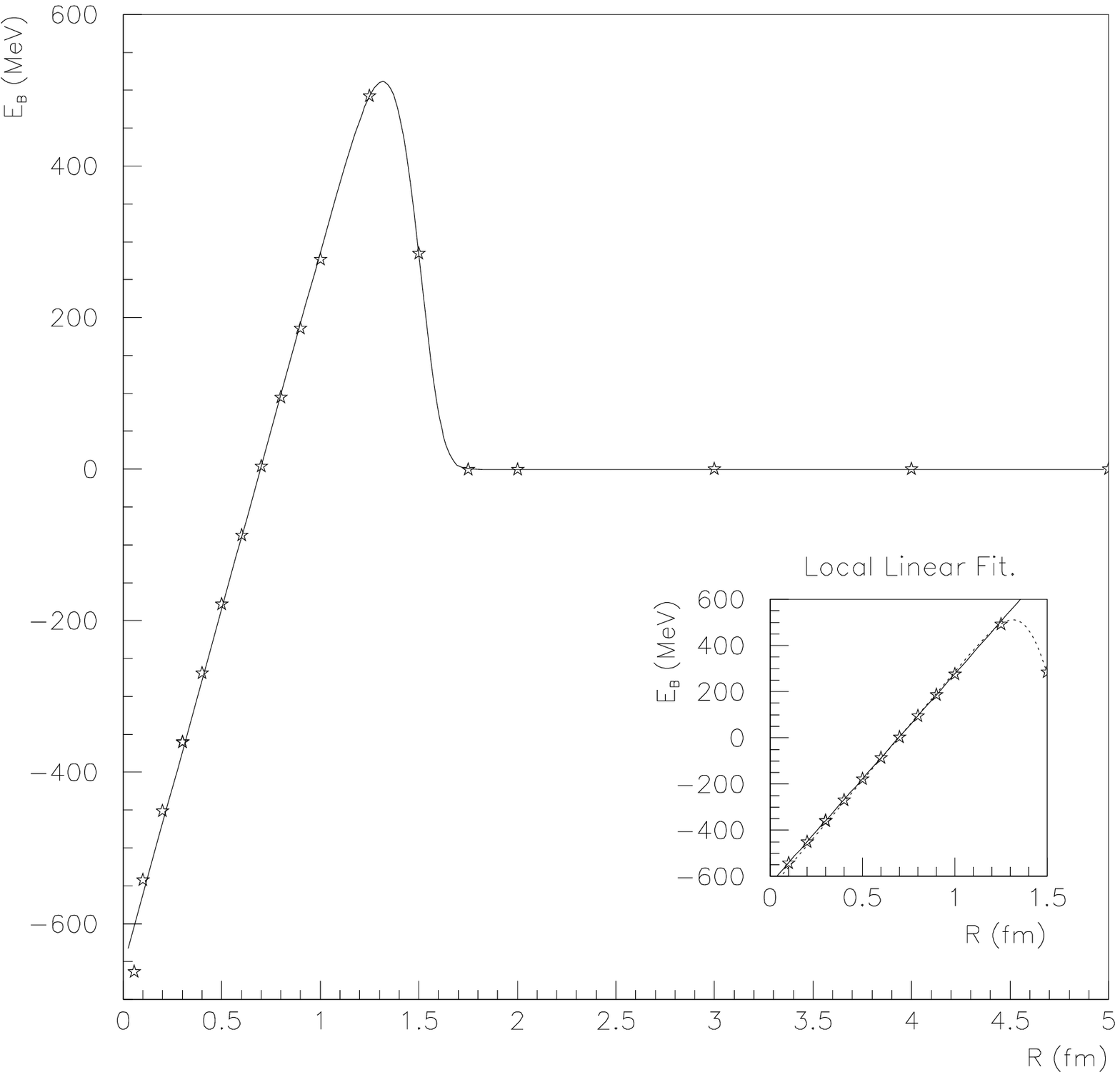}}}
\put(55,0 ){\mbox{\epsfxsize=6cm\epsffile{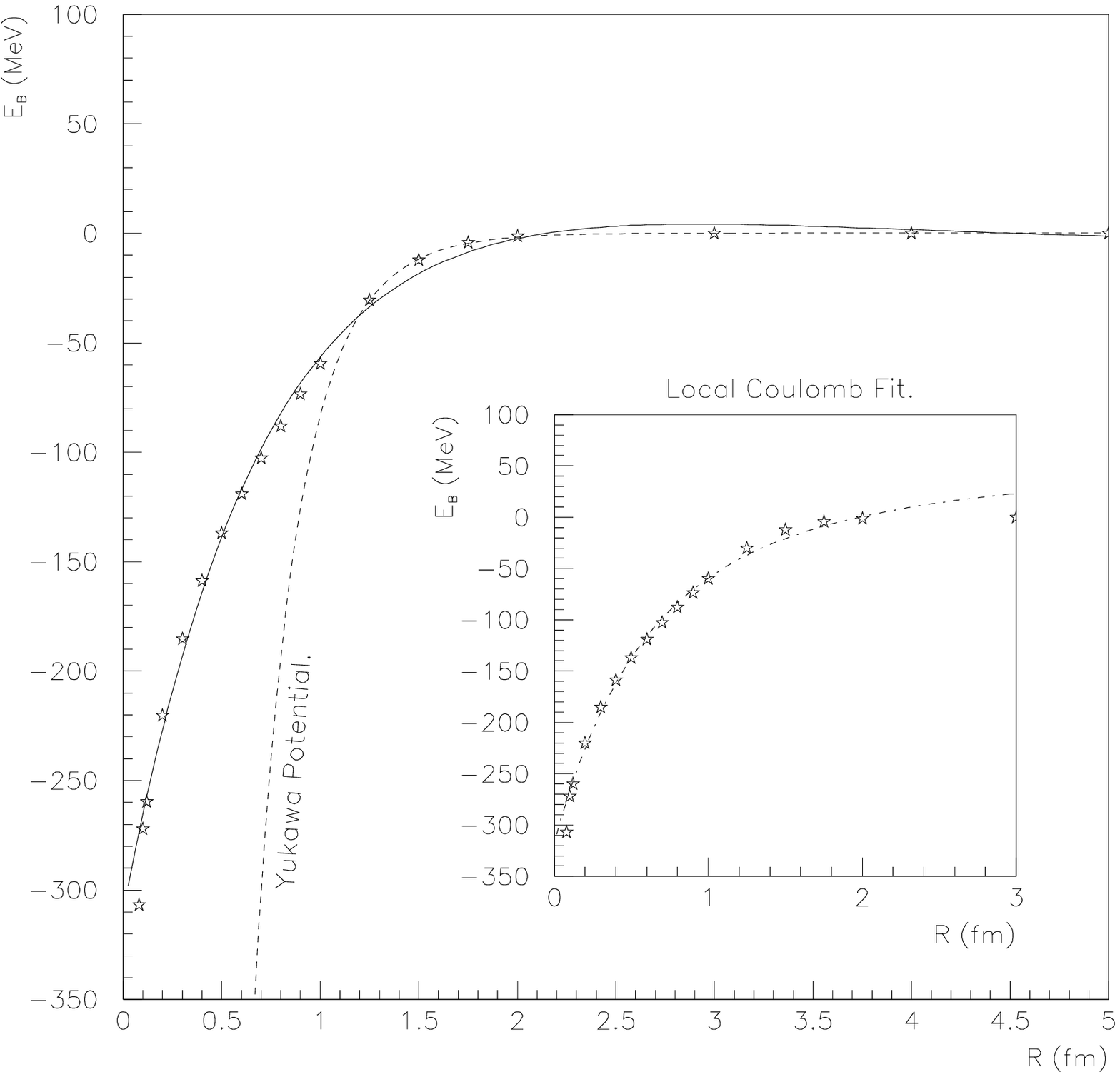}}}
\end{picture}}
\put(-4,10){
\parbox{11.5cm}{\footnotesize {\bf Fig.~2:}  Binding  energy as function
of heavy-quark separation  for the  old (a) and  new  (b) wave functions
with fixed-colour, and respective plots (c) and  (d) for the flux-bubble
model  with   moving-colour.  Shown   are  the  string-flip  (linear and
linear-plus-Coulomb) and flux-bubble (plus an $SU_c(3)$ composite model)
models for both wave functions,  and semi-relativistic plots for the old
wave function with fixed-colour.}}
\end{picture}
}
\end{center}
  
  Although  the $Q_2$ system is  far  removed from  its $H_2$,  hydrogen
molecule, cousin from a dynamical point  of view and the motivations for
achieving localization are quite  different, it would seem plausible  to
use a similar $ansatz$:\cite{kn:Schiff,kn:Bransden}
\begin{equation}
\Psi_{\alpha,\beta}=
\dspexp{-\beta^\alpha(r_{\bar q_1Q_1}^\alpha+r_{\bar q_2Q_2}^\alpha)}+
\dspexp{-\beta^\alpha(r_{\bar q_2Q_1}^\alpha+r_{\bar q_1Q_2}^\alpha)}\,,
\label{eq:psdo}
\end{equation}
where setting $\alpha=1$ yields  the $H_2$ case.   Now, if $\bar q_1Q_1$
and  $\bar  q_2Q_2$ represent two separate  mesons,  say, then the first
term represents the  internal  meson interactions while the  second term
represents  the external meson   interactions. Notice that  the external
interactions   shut  off as  the   separation,  $R\,$, between  the  two
heavy-quarks becomes large, which is the desired property.

  Fig.~1.b, shows  the results   for  linear,  linear-plus-coulomb,  and
flux-bubble potential models with fixed-colour, using the new $H_2$-like
variational wave function.\footnote{Also shown, are results for $\alpha$
fixed at 2 and 1.74$\,$.}  The contrast between  the results for the old
and new wave functions is quite dramatic!  The new wave function gives a
factor of ${\cal O}(27)$ increase in  well depth; deep  enough to bind a
heavy-quark   system     with a   reduced-mass    greater  than   ${\cal
O}(660)MeV$.\cite{kn:thesisb}

  Figs.~1.c and 1.d show the   results for the flub-bubble models   with
moving-colour for the old and  new wave function respectively.  Fig.~1.c
shows a  linear core  (with  slope  $\sigma$) with  a  repulsive barrier
followed by  an abrupt cut-off at 1.5$fm$.   This  suggests that minimal
energy  state is   a  system consisting  of two   mesons, one containing
heavy-quarks and  the other containing light-antiquarks.  Fig.~1.d shows
potential with a  very strong Coulomb-like ({\it i.e.}, $\alpha\sim{\cal
O}(125.5)MeV\,fm$) interior,   about $300MeV$ deep,   with a Yukawa-like
exterior.  This suggests a very tightly bound $Q_2$ system, however, the
$Q_2$-well  for the old   wave function is  much deeper,  suggesting the
aforementioned two-meson system is the  minimal energy state.  Of course
this is not physical,  and  suggests that phenomenological  models using
$SU_c(2)$, with moving-colour, will yield unsatisfactory results.
   
  Fig.~1.b, shows the  results   for an  $SU_c(3)$ composite   model,  a
``baryonic molecule,'' which assumes the two heavy-quarks are actually a
composite  of two light-quarks (also,  the orbiting antiquarks have been
replaced with quarks).     This model avoids dissociation  into  mesons,
since a  flux-tube  can  no longer  be  formed between   the (composite)
heavy-quarks, thus,  restoring the potential  to  its generic (original)
form.

  It  should  be  stressed  that the  $Q_2$  system  is  a toy-model,  a
pedagogical aid, used to gain better insight into  the complex nature of
nuclear interactions. Nuclear systems (at   low temperature) consist  of
equal mass quarks,  whereas, the $Q_2$ consists  of unequal mass quarks.
Therefore, what  is applicable for  $Q_2$ system may not  translate very
well to nuclear matter.  That said, the results herein, suggest that the
many-body wave function for nuclear matter should be of the form,
\begin{equation}
\Psi\sim{\rm Perm}\left|\,\dspexp{-(\beta r_{ij})^\alpha}\right|\,
\prod_{\mbox{\tiny colour}}|\Phi(r_{k})|\,.
\label{eq:ansatz}
\end{equation}
which is a product of   a  totally symmetric correlation wave   function
times a  totally antisymmetric Slater wave function.    A simple test of
this, {\it ansatz},  would be  to consider  $\Psi$ in  the context of  a
$q\bar q$ gas with fixed-$SU_c(2)$.

\section{Conclusions}

  The $Q_2$ system has proven to be a very useful aid for trying to sort
out the complexities of model  building for nuclear matter.  The details
of the mechanics, from wave functions to dynamics to practical computing
methods, of the   flux-bubble  model have been   throughly investigated.
However,   it remains to  be  see how applicable  this   work will be to
nuclear matter;   something that we  hope  to  investigate  in the  near
future.

  In closing, it appears that  the flux-bubble model  may prove not only
useful for modeling nuclear matter, but also useful for modeling mesonic
molecules as well.

\section*{Acknowledgments}

 This  work was supported by NSERC.   The  Monte Carlo calculations were
performed  on an 8 node DEC  5240 UNIX CPU  farm, using Berkeley Sockets
(TCP/IP).\cite{kn:Stevens} The computing facilities were provided by the
Carleton University's Dept.  of  Physics, OPAL, CRPP, and  T{\scriptsize
HEORY}.  We  would like to  thank M.A.~Doncheski, H.~Blundell, M.~Jones,
and J.S.~Wright for useful discussions.

\end{document}